\documentstyle[twocolumn,aps,epsf]{revtex}
\begin{document}
\draft
\title{Linear optical absorption spectra of mesoscopic 
structures in intense THz fields:\\ free particle properties}
\author
{Kristinn Johnsen$^{1,2}$ and Antti-Pekka Jauho$^1$}
\address{
$^1$Mikroelektronik Centret, Technical University of Denmark, Bldg 345east\\
DK-2800 Lyngby, Denmark\\
$^2$Dept. of Mathematics, University of California, Santa Barbara, CA93106, USA
\medskip\\
\date{Submitted to Physical Review B on April 3, 1997}
\parbox{14cm}{\rm
We theoretically study the effect of THz radiation on the 
linear optical absorption spectra 
of semiconductor structures. A general theoretical framework, 
based on non-equilibrium
Green functions, is formulated, and applied to the calculation of
linear optical absorption spectrum for several non-equilibrium
mesoscopic structures. We show that a blue-shift occurs and sidebands
appear in bulk-like structures, i.e., the dynamical Franz-Keldysh effect
[A.-P. Jauho and K. Johnsen, Phys. Rev. Lett. {\bf 76}, 4576 (1996)]. 
An analytic calculation leads to the prediction that in the case of 
superlattices distinct stable steps appear in the absorption 
spectrum when conditions for dynamical localization are  met.
\smallskip\\
PACS numbers: 71.10.-w,78.20.Bh,78.20.Jq,78.47.+p
\smallskip\\}
}
\maketitle
\narrowtext
\section{Introduction}

Light absorption can be described in terms of
a process where polarization is 
induced in the medium.
To linear order in the electric
field component of the traversing light, ${\cal E}$,
the induced polarizability, ${\cal P}$,
can be expressed in terms of the dielectric susceptibility $\chi$ as
$$
{\cal P}(t) = \int_{-\infty}^t{\rm d}t'\,\chi (t,t'){\cal E}(t').
$$
If the absorbing medium is in a stationary state, the
susceptibility depends only on the difference of its time arguments, i.e.,
$\chi (t,t') = \chi (t-t')$. Under these conditions
Maxwell's equation for ${\cal E}$
is an algebraic equation in frequency space and one finds that the absorption
is proportional to the imaginary part of $\chi (\omega )$.
However, under non-equilibrium conditions, which are the
topic of the present study, the susceptibility is
a two-time function, and Maxwell's
equation remains an integral equation even in the
frequency domain.
Further progress hinges upon two steps: first, one has to develop
methods to calculate the non-equilibrium susceptibility function
and second, 
one has to specify what sort
of light wave or pulse is used in the absorption experiment.
The present work addresses both of these problems.
As far as the time-dependence of the probe pulse is concerned,
two specific situations are examined.
First, consider an undoped semiconductor placed in an intense 
THz field; we assume that the THz field is not able to induce polarization, 
i.e., no carriers are excited in the conduction band. Such a system is in  
a non-equilibrium state, i.e., the susceptibility is a two time function.
Properties of such systems have been investigated experimentally using
the Free Electron Laser (FEL) as a source for intense THz 
fields\cite{CER95};  many interesting properties 
have been discovered, and others predicted,
 such as photon-assisted tunneling \cite{GUI93},
dynamical localization and absolute negative conductivity \cite{KEA95},
ac Stark effect\cite{HOL94}, 
dynamical Franz-Keldysh effect
and formation of sidebands\cite{YAC68,JAU96,KON97,BIR95}.
A second example consists of ultra-fast transients. 
Consider an undoped semiconductor structure subject to an external 
intense static field. At some time instant a population of carriers is
pumped into the conduction band. These mobile charges will rearrange
themselves so as to 
screen the external field. While the screening is building up 
the susceptibility of the system is a two time function. 
Using femtosecond laser techniques, which are able to probe
the time scales in which screening is building up, experiments
investigating the non-equilibrium properties of such systems have been
performed\cite{HEE93,CAM96}.

Bandgap engineering techniques of semiconductor compounds, 
such as Molecular Beam Epitaxy (MBE), allow
spatial modulation of the bandgap down to atomic resolution. 
It is possible to break the translational symmetries
of bulk crystals, induce new ones and reduce
the degrees of freedom with these techniques.
Construction of systems which are effectively two dimensional (2D) and even
one dimensional (1D) with regard to electron mobility and optical
properties is today a standard procedure. Another example
of such man-made structures are superlattices (SL), i.e.,
an engineered periodic potential in the growth direction of the  
sample. The interplay between the mesoscopic properties and
the dynamical properties can lead to many interesting phenomena
such as absolute negative resistance for the transport 
properties\cite{WAC97} and the rich features in the optical properties
which are the subject of the present work. 

The purpose of this paper is to present a
theoretical study of light absorption in  mesoscopic systems 
subject to intense THz [far infrared (FIR)] fields.
We consider  undoped  systems, which implies that there are no carriers
in the conduction band.  Thus, the near infrared (NIR) inter-band 
absorption is the dominant absorption process. 
The non-equilibrium nature of the system necessitates the
use of special theoretical tools; we have
chosen to apply the non-equilibrium Green function 
techniques\cite{HAU96,POT96}. 
In particular, this method allows us to treat the intense FIR field
non-perturbatively, and defines a framework in which screening
can be treated systematically. 
Our analysis consists of the following steps.
Starting from a two-band Hamiltonian we derive a 
formal 
expression for the inter-band susceptibility in terms of non-equilibrium
Green functions. Next, we use the general
expression to derive the NIR absorption spectrum
for non-interacting particles 
(Coulomb interactions will be discussed in a subsequent paper,
see also below). Finally, we give explicit results for a number
of special cases (3-, 2-, 1-d systems and superlattices), and
discuss the physical implications.

This paper is organized as follows. In section \ref{susc_sec} we derive 
a general expression for the two-time
dielectric inter-band susceptibility. Section \ref{abs_sec} 
relates the susceptibility to the measured absorption, considering both
continuous wave measurements as well as short white light pulses.
The single-particle Green functions and the corresponding spectral
functions, which determine the susceptibility, are defined
in section \ref{gre_sec} and related to the generalized density of 
states\cite{JAU96} (GDOS), which, in turn, is shown to determine the 
optical absorption.
Section \ref{bulk_sec} considers bulk-like
systems, and we obtain analytic results for the GDOS,
which is analyzed in some detail in terms of the side-band picture.
The properties of light absorption in superlattices are treated in section
\ref{sup_sec}. Specific attention is paid to conditions where
dynamical localization occurs,
and we show how it affects the absorption spectrum. Finally,
in section \ref{con_sec} we have some concluding remarks.
\section{The dielectric inter-band susceptibility}
\label{susc_sec}

We shall now derive an expression for the dielectric inter-band 
susceptibility 
using non-equilibrium Green functions. 
The microscopic operator describing inter-band 
polarization is
\begin{equation}
\vec{P}(t) = \sum_{k}\vec d_k \left[a_k^\dagger (t)b_k(t)+b_k^\dagger (t)a_k(t)
\right]\;.
\label{pol}
\end{equation}
Here $\vec d_k$ is the dipole matrix-element, $a_k^\dagger (t)$ [$a_k(t)$]
are the conduction band electron creation 
[annihilation] operators and $b_k^\dagger (t)$ 
[$b_k(t)$] are the valence band creation [annihilation] 
operators. The linearized Hamiltonian associated
with a polarization $\vec{P}(t)$ induced by the external field ${\cal E}(t)$ is
$H_P(t) = - \vec{P}(t)\cdot {\cal E}(t)$.
Linear response theory now yields the Cartesian $l$-component
of the
induced inter-band polarization
due to a weak external field ${\cal E}$:
\begin{equation}
{\cal P}_l(t) = -{i\over\hbar}\int_{-\infty}^\infty {\rm d}t'\,\theta (t-t')
\langle [ \vec{P}(t'),P_l(t)]\rangle\cdot {\cal E}(t').
\label{is5}
\end{equation}
The retarded susceptibility tensor can be identified from Eq.(\ref{is5}):
\begin{equation}
\chi_{lm}^r(t,t') = -{i\over\hbar}\theta (t-t')
\langle [ P_m(t'),P_l(t)]\rangle.
\end{equation}
Following the standard line of attack in non-equilibrium theory \cite{HAU96},
we first consider the causal (time-ordered) response function:
\begin{equation}
\chi_{lm}^c(t,t') =-{i\over\hbar}\langle {\rm T}\{ P_m(t')P_l(t)\}\rangle
\end{equation}
where ${\rm T}$ is the time-ordering operator. 
In non-equilibrium, the causal response function is replaced by its
contour ordered counterpart:
$\chi_{lm}^c(t,t')\to\chi_{lm}^c(\tau,\tau')$,
where the complex-time variables $\tau,\tau'$ reside on
the Keldysh contour. Finally we obtain the retarded
tensor by an analytic continuation
using the Langreth rules\cite{HAU96Table41}.

We use Eq.(\ref{pol}) to write the susceptibility as 
\begin{eqnarray}\label{chifull}
\chi_{lm}^c(\tau,\tau') 
&=& -{i\over\hbar}\sum_{q\,k}d_l(k)d_m(q) \\
&\times&[\langle{\rm T}_c\{a_q^\dagger (\tau')b_q(\tau')
a_k^\dagger (\tau)b_k(\tau)\}\rangle
\nonumber \\
&&+\langle{\rm T}_c\{a_q^\dagger (\tau')b_q(\tau')
b_k^\dagger (\tau)a_k(\tau)\}\rangle
\nonumber \\ &&
+\langle{\rm T}_c\{b_q^\dagger (\tau')a_q(\tau')
a_k^\dagger (\tau)b_k(\tau)\}\rangle
\nonumber \\
&&+\langle{\rm T}_c\{b_q^\dagger (\tau')a_q(\tau')
b_k^\dagger (\tau)a_k(\tau)\}\rangle
]\;. \nonumber
\end{eqnarray}
where ${\rm T}_c$ is the contour-ordering operator.
In equilibrium, the  two-particle 
correlation functions occurring in Eq.(\ref{chifull}) would be found via the
Bethe-Salpeter equation\cite{HAU84}.
In what follows, however, we shall consider the noninteracting limit
of Eq.(\ref{chifull}).
This approach is motivated by the
following considerations.
The noninteracting limit will allow significant analytic progress,
and the results, which we believe are interesting in their
own right,  will form the basis for
any subsequent interacting theories.
Secondly, experimentally it is known that in undoped semiconductor
quantum wells excitons are quenched if the system is subject to intense
FIR\cite{NOR97}, thus implying that for the present
situation Coulomb interactions
may not play an equally dominant role as is the case in
equilibrium situations.
A quantitative assessment requires a non-equilibrium theory for the
two-particle Green functions and will be a topic of our future
work.  For  noninteracting particles we can  use Wick's theorem 
to factorize the two-particle correlation functions.
Thus, the non-equilibrium susceptibility can be expressed
in terms of single-particle Green functions.
The following Green functions are needed:
\begin{eqnarray}
g_c(k,\tau;q,\tau') = -i\langle{\rm T}_c\{a_k(\tau)a_q^\dagger (\tau')\}
\rangle \\
g_v(k,\tau;q,\tau') = -i\langle{\rm T}_c\{b_k(\tau)b_q^\dagger (\tau')\}
\rangle \\
g_{ab}(k,\tau;q,\tau') = -i\langle{\rm T}_c\{a_k(\tau)b_q^\dagger (\tau')\}
\rangle
\end{eqnarray}
and
\begin{equation}
g_{ba}(k,\tau;q,\tau') = -i\langle{\rm T}_c\{b_k(\tau)a_q^\dagger (\tau')\}\rangle.
\end{equation}
We assume that the frequency, $\Omega$, of the FIR
field is such that $\hbar\Omega\ll\epsilon_g$. 
In typical experiments on III-V systems $\epsilon_g$ is of the
order of eV, while $\hbar\Omega$ is a few meV, so this
condition is satisfied.
Consequently, inter-band transitions due to the 
perturbing field can be ignored, and
the Green functions related to  Zener-effect, i.e.,
$g_{ab}(k,\tau;q,\tau')$ and $g_{ba}(k,\tau;q,\tau')$, 
are neglected from this on.
The first order non-equilibrium susceptibility reads thus
\begin{eqnarray}
\chi_{lm}^c(\tau,\tau') &=&
-{i\over\hbar}\sum_{q\,k}d_l(k)d_m(q)[
g_c(k,\tau;q,\tau')g_v(q,\tau';k,\tau)\nonumber\\
&&+g_v(k,\tau;q,\tau')g_c(q,\tau';k,\tau)
].
\end{eqnarray}
The analytic continuation to real times is performed with 
the Langreth rules \cite{HAU96Table41}, which state that
if on contour 
\begin{equation}
C(\tau ,\tau') = A(\tau ,\tau')B(\tau' ,\tau)\;,
\nonumber
\end{equation}
then the retarded function on the real time axis is
\begin{equation}
C^r(t,t') =  A^<(t ,t')B^a(t',t) + A^r(t,t')B^<(t',t).
\nonumber
\end{equation}
We thus have
\begin{eqnarray}
\chi_{lm}^r(t,t') &=& -{i\over\hbar}\sum_{k}d_l(k)d_m(k)[
g_c^<(k,t,t')g_v^a(k,t',t)\nonumber\\
&&+g_c^r(k,t,t')g_v^<(k,t',t)
+g_v^<(k,t,t')g_c^a(k,t',t) \nonumber \\
&&+g_v^r(k,t,t')g_c^<(k,t',t) ],
\end{eqnarray}
with
\begin{eqnarray}
g^<(t,t') &=& i\langle c^\dagger (t')c(t)\rangle ,\\
g^a(t,t') &=& i\theta (t'-t)\langle\{c(t),c^\dagger (t')\}\rangle ,\\
g^r(t,t') &=& -i\theta (t-t')\langle\{c(t),c^\dagger (t')\}\rangle.
\end{eqnarray}
We recall the following relations:
\begin{eqnarray}
\left[g^<(t,t')\right]^* &=& - g^<(t',t) \nonumber\\
\left[g^a(t,t')\right]^* &=& g^r(t',t) \\
\left[g^r(t,t')\right]^* &=& g^a(t',t)\nonumber
\end{eqnarray}
For certain applications, e.g. Section
\ref{cw_sec}, it is convenient to introduce the center of mass variables
$T = (t'+t)/2$ and $\tau = t-t'$ 
\cite{notation}.
In terms of these variables the symmetry
relations of the Green functions are:
\begin{eqnarray}
\left[g^<(T,\tau )\right]^* &=& - g^<(T,-\tau ) \nonumber\\
\left[g^a(T,\tau )\right]^* &=& g^r(T,-\tau ) \\
\left[g^r(T,\tau )\right]^* &=& g^a(T,-\tau )\nonumber
\end{eqnarray}
The retarded susceptibility expressed in center of mass coordinates
is
\begin{eqnarray}
\label{chicm}
\chi_{lm}^r(T,\tau) &=& -{i\over\hbar}\sum_{k}d_l(k)d_m(k)\nonumber \\
&\quad&\times\{[ g_c^<(k,T,\tau )g_v^a(k,T,-\tau ) \nonumber \\
&&\quad + g_v^<(k,T,\tau )g_c^a(k,T,-\tau ) ] - h.c.\}.
\end{eqnarray}
Note that in equilibrium $\chi_{lm}^r(T,\tau) = \chi_{lm}^r(\tau)$.
As shown  in Section \ref{cw_sec}, the relevant quantity for
continuous wave measurements at frequency $\omega_l$ is
\begin{equation}
{\rm Im}\chi_{lm}^r(T,\omega_l ) = {\rm Im}
\left\{\int_{-\infty}^\infty{\rm d}\tau\,
e^{i\omega_l\tau}\chi_{lm}^r(T,\tau) \right\}
\end{equation}
to first order in $\Omega /\omega_l$ (here $\Omega$ is the FIR
frequency). Now, 
$\chi_{lm}^r(T,\tau)$ is a real quantity as is evident from (\ref{chicm}). 
Using the properties of the Fourier transform 
we obtain 
\begin{eqnarray}
\chi_{lm}^r(T,\omega_l ) &=& 
-{i\over\hbar}\sum_{k}d_l(k)d_m(k)
\int_{-\infty}^\infty {{\rm d}\omega\over 2\pi}\Big\{
g_c^<(k,T,\omega )  \nonumber \\ 
&&
\times [ g_v^a(k,T,\omega-\omega_l ) + g_v^r(k,T,\omega + \omega_l )]
 \nonumber \\
&&+
g_v^<(k,T,\omega )\nonumber \\ 
&&\times 
[g_c^a(k,T,\omega-\omega_l )
+ g_c^r(k,T,\omega+ \omega_l )]\Big\}.\nonumber\\
\end{eqnarray}
Since $\chi_{lm}^r(T,\tau )$ is real, the 
imaginary part of its Fourier transform is obtained through
\begin{equation}
\nonumber
{\rm Im}\chi_{lm}^r(T,\omega_l ) =
{1\over 2i}[\chi_{lm}^r(T,\omega_l ) - \chi_{lm}^r(T,-\omega_l )].
\end{equation}
We can therefore write, in terms of the spectral functions
\begin{equation}
\nonumber
A_c(k,T,\omega ) = i[g_c^r(k,T,\omega )-g_c^a(k,T,\omega )]
\end{equation}
and
\begin{equation}
\nonumber
A_v(k,T,\omega ) = i[g_v^r(k,T,\omega )-g_v^a(k,T,\omega )],
\nonumber
\end{equation}
that
\begin{eqnarray}\label{ImchiT}
{\rm Im}\chi_{lm}^r(T,\omega_l )
&=&{i\over 2\hbar }
\sum_{k}d_l(k)d_m(k)
\int_{-\infty}^\infty {{\rm d}\omega\over 2\pi}\Big\{g_c^<(k,T,\omega ) 
\nonumber \\
&&\times
[ A_v(k,T,\omega-\omega_l)-A_v(k,T,\omega+\omega_l) ] \nonumber \\
&&+
g_v^<(k,T,\omega ) \nonumber \\ &&\times
[ A_c(k,T,\omega-\omega_l)-A_c(k,T,\omega+\omega_l)] \Big\}.\nonumber
\\
\end{eqnarray}
The lesser functions can be expressed in the form\cite{HAU96}
\begin{equation}
\nonumber
g_a^<(k,T,\omega ) = if_a(k,T,\omega ) A_a(k,T,\omega )
\end{equation}
where $f_a(k,T,\omega )$ is a generalized
particle distribution for particles of species
$a$, and $A_a(k,T,\omega )$ is the corresponding spectral function.
In accordance with our assumption about no
FIR field induced inter-band transitions, we can set
$f_c(k,T,\omega )=0$ (zero occupation of conduction
band), and that $f_v(k,T,\omega )=1$ (all valence states are occupied).
In the general case, e.g., when considering nonlinear effects
in the {\it probing} light field,
one would have to find $f_a(k,T,\omega )$ via, say, 
a Monte Carlo solution of semiconductor
Bloch equations \cite{HAU93,KUH92}
or by a direct integration of quantum kinetic
equations for
$g_a^<(k,T,\omega )$ \cite{HAU96}.
With these assumptions the susceptibility reduces to
\begin{eqnarray}\label{convolution}
{\rm Im}\chi_{lm}^r(T,\omega_l )
&=& {-1\over 2\hbar }\sum_{k}d_l(k)d_m(k)
\int_{-\infty}^\infty {{\rm d}\omega\over 2\pi}
 A_v(k,T,\omega )\nonumber\\
&\quad&\times \{ A_c(k,T,\omega-\omega_l)-A_c(k,T,\omega+\omega_l) \} 
\nonumber \\
&=&
{1\over 2\hbar }\sum_{k}d_l(k)d_m(k)\nonumber\\
&\quad&
\times\int_{-\infty}^\infty {{\rm d}\omega\over 2\pi}
A_v(k,T,\omega )A_c(k,T,\omega+\omega_l). \nonumber\\
\end{eqnarray}
The second equality comes about because we do not consider
overlapping bands. Eq.(\ref{convolution}), which is the central
result of this section, expresses the fact that the non-equilibrium
inter-band susceptibility function can be calculated from a
{\it joint spectral function}, which is a convolution of the
individual band spectral function.  A similar result is known
from high-field quantum transport theory \cite{HAU96Chap11}: there the
field-dependent scattering rate is expressed as a joint spectral
function for the initial and final states.

In order to make a connection to the equilibrium case,
we recall the exact identity $g^<_{\rm eq}(k,\omega)
=ia_{\rm eq}(k,\omega)n(\omega)$ ($n(\omega)$ is
the Fermi function), and obtain from Eq.(\ref{ImchiT})
\cite{HAU84,SCH89}
\begin{eqnarray}
{\rm Im}\chi_{lm}^r(\omega_l )
&=&-{1\over 2\hbar}
\sum_{k}d_l(k)d_m(k)
\nonumber\\
&\quad&\times\int_{-\infty}^\infty {{\rm d}\omega\over 2\pi}
\Big\{ n_c(\omega )a_{c,{\rm eq}}(k,\omega ) \nonumber \\
&\quad&\times [a_{v,{\rm eq}}(k,\omega-\omega_l)-
a_{v,{\rm eq}}(k,\omega+\omega_l)] \nonumber \\
&\quad&+ n_v(\omega )a_{v,{\rm eq}}(k,\omega ) \nonumber \\
&\quad& \times [ a_{c,{\rm eq}}(k,\omega-\omega_l)-
a_{c,{\rm eq}}(k,\omega+\omega_l)] \Big\}.\nonumber\\
\end{eqnarray}
Here $n_c(\omega )$ is the conduction band 
electron occupation function,
and $n_v(\omega )$ is corresponding function for the valence band electrons.
\section{Absorption coefficient in terms of the time dependent
dielectric susceptibility}
\label{abs_sec}

The dielectric susceptibility $\chi$ links the 
induced polarization
${\cal P}$ to the field ${\cal E}$ via
\begin{equation}
{\cal P}(t) = \int_{-\infty}^t{\rm d}t'\,\chi (t,t'){\cal E}(t').
\label{ad1}
\end{equation}
The wave equation for light is then
\begin{equation}\label{waveq}
\nabla^2{\cal E}(t) - {1\over c^2}{\partial^2 {\cal D}(t)\over\partial t^2} 
= 0
\end{equation}
where
$
{\cal D}(t) = {\cal E}(t) + 4\pi {\cal P}(t)$.
The absorption coefficient $\alpha (\omega )$ is defined as 
the inverse of the length which light has to traverse 
in the medium at frequency $\omega$ in 
order for the intensity of the light
to decrease by a factor of $1/e$.  In equilibrium 
$
{\cal D}(\omega ) = [1+4\pi\chi (\omega)]{\cal E}(\omega )
= \epsilon (\omega){\cal E}(\omega )
$
and the  
absorption coefficient \cite{HAU93} becomes
\begin{equation}\label{alphastat}
\alpha (\omega ) = 4\pi\omega{{\rm Im}\chi (\omega )\over cn (\omega )}.
\end{equation}
Here $n^2(\omega)= \frac{1}{2}
[{\rm Re}\epsilon(\omega)+|\epsilon(\omega)|]$ is the 
refraction coefficient which
usually depends only weakly 
on $\omega$. 
In non-equilibrium this analysis must be generalized, and
we consider two special cases:
(i) monochromatic continuous wave 
measurements, and (ii) white light short pulse measurements.
\subsection{Continuous wave measurements}
\label{cw_sec}

Consider a system out 
of equilibrium which is probed by a light field (which is assumed
to be weak)
of frequency $\omega_l$:
\begin{equation}
{\cal E}(r,t) = {\cal E}_0 \exp [i(r\cdot k-\omega_l t)].
\end{equation}
The polarization can then be expressed as
\begin{equation}\label{pol1}
{\cal P}(t) = {\cal E}(r,t)
\int_{-\infty}^{\infty} dt'
e^{i\omega_l(t-t')}\chi^r(t,t')\;.
\end{equation}
This form is suggestive: it is advantageous to
express $\chi^r$ in terms of the center-of-mass and difference
coordinates, $\chi^r(t,t')\to{\tilde\chi}^r({1\over 2}(t+t'),t-t')$.
The characteristic time-scale for the center-of-mass time
is set by the ``slow'' frequency $\Omega$, while the difference-time
varies on the scale of the ``fast'' frequency $\omega_l$.  We thus
gradient-expand:
${\tilde\chi}^r({1\over 2}(t+t'),t-t')\simeq{\tilde\chi}^r(t,t-t')+
{1\over 2}(t'-t) 
\tilde\chi^{r^\prime}(t,t-t') + \cdots$,
where the prime indicates differentiation with respect to the slow
temporal variable. Substitution in (\ref{pol1}) then yields
(we introduce a new variable $\tau\equiv t-t'$)
\begin{eqnarray}\label{gradexp}
{\cal P}(t)&=&{\cal E}(r,t)\int d\tau e^{i\omega_l\tau}[
{\tilde\chi}^r(t,\tau) + (-{1\over 2}\tau)\tilde\chi^{r^\prime}(t,\tau)
+\cdots]\nonumber\\
&=&{\cal E}(r,t)[{\tilde\chi}^r(t,\omega_l) + {\partial\over\partial\omega_l}
{\partial\over\partial t} {i\over 2}{\tilde\chi}^r(t,\omega_l)+\cdots]\nonumber\\
&=&{\cal E}(r,t) \exp\left[{i\over 2}{\partial^2\over\partial t
\partial\omega_l}\right]
{\tilde\chi}^r(t,\omega_l)\;.
\end{eqnarray}
Eq.(\ref{gradexp}) can now be used in the Maxwell equation; note however
that upon Fourier-transforming the dominant frequency comes from ${\cal E}(t)$
and we can keep $t$ in $\chi(t,\omega_l)$ fixed.  The slow time-variation
will from this on be indicated by $T$.
Proceeding as in deriving the static result (\ref{alphastat}),
we identify the {\it time-dependent} absorption
coefficient
\begin{equation}
\alpha_T(\omega ) =
4\pi\omega{{\rm Im}{\tilde\chi}^r (T,\omega)\over cn_T(\omega )} + 
{\cal O}(\Omega /\omega ).
\end{equation}
If the driving force is periodic in $T$ (the harmonic time-dependence
due to a FEL-laser is an important special case), then the average
absorption coefficient is
\begin{eqnarray}
\bar\alpha (\omega ) &=& {1\over T_{\rm period}}\int_{\rm period}{\rm d}T\,
\alpha_T(\omega ) \\
&=& {1\over T_{\rm period}}\int_{\rm period}{\rm d}T\,
4\pi\omega{{\rm Im}{\tilde\chi}^r (T,\omega)\over cn_T(\omega )} \nonumber
\end{eqnarray}
to {\it all} orders in $\Omega /\omega$. We stress that here ${\tilde\chi}^r (T,\omega)$
is Fourier transformed with respect to the difference variable $\tau$.
Below we shall represent numerical examples for the generalized
absorption coefficient.
\subsection{Short white light pulse measurements}
\label{sl_sec}

Consider now an instantaneous measurement performed on
a non-equilibrium system: at some specific
time $t=t_m$ the system  is probed
with a weak pulse whose duration is short compared to the characteristic
dynamics of the system. 
We approximate the pulse with delta function 
in time:
\begin{equation}\label{deltapulse}
{\cal E}(r,t) 
= {\cal E}_0e^{ir\cdot k} \delta (t-t_m).
\end{equation}
In principle, construction of such a pulse would take infinite energy due to its
time dependence. The pulse is therefore hardly ``weak''. When we refer to the 
pulse as weak we assume that ${\cal E}_0 \ll 1$, that is the intensity of the
light is small at all frequencies. 
Using (\ref{deltapulse}) in the Maxwell equation 
yields dispersion relation
\begin{equation}
k^2 = {\omega^2\over c^2}[1+4\pi\chi^r (\omega,t_m)].
\end{equation}
This dispersion relation looks quite
similar to the one obtained in the 
previous section. The difference is that here $\chi^r (\omega,t_m)$ is Fourier
transformed with respect to $t'$, {\em not} the difference variable 
$\tau$. In the present case we obtain the time dependent
absorption coefficient
\begin{equation}
\alpha_t(\omega ) = 4\pi\omega{{\rm Im}\chi^r (\omega_l,t)\over cn_t(\omega )}.
\end{equation}
For examples of experiments which probe systems in this manner 
see, e.g., Refs. \cite{HEE93,CAM96}.

\subsection{Differential Transmission Spectrum}
\label{dts_sec}

Consider a sample of thickness $L$;
then the ratio of the intensity transmitted through the sample with its initial
intensity is $T(\omega )= \exp [-\bar\alpha (\omega )L]$ where 
$\bar\alpha (\omega )$
is the absorption coefficient of the sample. 
Experimental setups for 
measuring the change in absorption due to externally controlled perturbations
commonly measure the differential transmission spectrum (DTS) defined 
by\cite{HAU93}
\begin{equation}
DTS( \omega )= {T(\omega )-T_0(\omega )\over T_0(\omega )}.
\end{equation}
Here $T(\omega )$ is the transmission with the perturbation present and
$T_0(\omega )$ is the transmission through the unperturbed sample.
Below we give examples of $DTS(\omega)$ in non-equilibrium situations.
\section{Single particle Green functions and the spectral functions}
\label{gre_sec}

In this section we determine the single-particle 
Green functions and their associated spectral
functions. We show that, under conditions
specified below, that the convolutions of the spectral functions,
encountered in Section \ref{susc_sec},
result in effective single band spectral functions.

\subsection{The single particle Green functions}

Let $\vec A$ be the vector potential which defines the  
FIR field. Considering harmonic, translationally
invariant external fields we choose
\begin{equation}
\vec A (t) = -\vec E{\sin (\Omega t)\over\Omega},
\label{vecpot}
\end{equation}
which represents the physical uniform 
electric field $\vec E\cos (\Omega t)$.
The Hamiltonian describing the two bands is then
\begin{equation}
H = \sum_{\vec k}\left\{
 \epsilon_c[\vec k-e\vec A(t)]a_{\vec k}^\dagger a_{\vec k}
+\epsilon_v[\vec k-e\vec A(t)]b_{\vec k}^\dagger b_{\vec k}\right\}.
\end{equation}
The Dyson equation for the retarded/advanced free-particle Green function is
\begin{equation}
\left( i\hbar{\partial\over\partial t} -
\epsilon_\alpha [\vec k-e\vec A(t)]\right)
g_\alpha^{r/a} (\vec k,t,t') = \delta (t-t')\;,
\end{equation}
where $\alpha\in \{c,v\}$ 
is the band index. This equation is
readily integrated with the solutions
\begin{equation}
g_\alpha^{r/a}(\vec k,t,t') = \mp{i\over\hbar}\theta (\pm t\mp t') 
\exp\Big\{
-{i\over\hbar}\int_{t'}^t{\rm d}s\epsilon_\alpha [\vec k-e\vec A(s)]
\Big\},
\end{equation}
and the spectral function  $A=i(g^r-g^a)$ becomes
\begin{equation}
A_\alpha (\vec k,t,t')
={1\over\hbar}
\exp\Big\{
-{i\over\hbar}\int_{t'}^t{\rm d}s\epsilon_\alpha [\vec k-e\vec A(s)]
\Big\}.
\end{equation}
\subsection{Convolution of the spectral functions}
\label{conv_sec}

According to Section \ref{susc_sec} 
the susceptibility is obtained through the trace of a convolution of the
spectral functions. We shall now show that within the present model the
convolution of spectral functions results in a new effective single-band
spectral function. 

In terms of the center of mass variables, $\tau = t-t'$ and $T = (t+t')/2$,
we write the spectral functions as
\begin{eqnarray}
A_\alpha (\vec k,T,\omega)&=& {1\over\hbar}
\int_{-\infty}^\infty{\rm d}\tau e^{i\omega\tau} \\ &&
\times\exp\Big\{
-{i\over\hbar}
\int_{T-\tau/2}^{T+\tau/2}{\rm d}s\epsilon_\alpha [\vec k-e\vec A(s)]
\Big\}.\nonumber
\end{eqnarray}
Then the convolution,
\begin{equation}\label{convolution1}
b(\vec k,T,\omega_l ) =\hbar
\int_{-\infty}^\infty{{\rm d}\omega\over 2\pi}
A_e(\vec k,T,\omega)A_v(\vec k,T,\omega-\omega_l),
\end{equation}
of the two spectral functions becomes
\begin{eqnarray}
b(\vec k,T,\omega_l ) &=& {1\over\hbar}
\int_{-\infty}^\infty{\rm d}\tau e^{i\omega_l\tau}
\exp\Big\{
-{i\over\hbar}
\int_{T-\tau/2}^{T+\tau/2}{\rm d}s
\\&&\times
\big( \epsilon_c [\vec k-{e}\vec A(s)] -\epsilon_v [\vec k-e\vec A(s)]\big)
\Big\}.\nonumber
\end{eqnarray}
In the case of parabolic bands we have
\begin{equation}
\epsilon_c [\vec k] = {\hbar^2k^2\over 2m_e}\quad , \quad
\epsilon_v [\vec k] = -{\hbar^2k^2\over 2m_h}-\epsilon_g.
\end{equation}
Here $m_e$ is the electron mass, $m_h$ is the positive hole mass and
$\epsilon_g$ is the band gap. We define a single effective band for the
system:
\begin{equation}
\epsilon_{{\rm eff}}[\vec k] \equiv \epsilon_c [\vec k] - \epsilon_v [\vec k]
= {\hbar^2k^2\over 2m_{{\rm eff}}}+\epsilon_g,
\end{equation}
where
$
m_{{\rm eff}} = {m_em_h/( m_e+m_h)}
$
is the effective reduced  mass. Thus, the effective band is parabolic like
the original bands but with their reduced mass.
It is therefore evident that the convolution,
writing $A_{{\rm eff}}(\vec k,T,\omega_l ) = b(\vec k,T,\omega_l )$, 
is a spectral function for a parabolic band,
\begin{eqnarray}
A_{{\rm eff}} (\vec k,T,\omega)&=& {1\over\hbar}
\int_{-\infty}^\infty{\rm d}\tau e^{i\omega\tau} \\ &&
\times\exp\{
-{i\over\hbar}
\int_{T-\tau/2}^{T+\tau/2}{\rm d}s\epsilon_{{\rm eff}}
[\vec k-e\vec A(s)]
\}.\nonumber
\end{eqnarray}
In the case of tight-binding minibands for a type I superlattice
(with period $a$) we write the
bands as
\begin{eqnarray}
\epsilon_c[\vec k] &=& {1\over2}\lambda_c\cos (ak_\parallel )
+{\hbar^2k_\perp^2\over 2m_e}
\\
\epsilon_v[\vec k] &=& -{1\over2}\lambda_v\cos (ak_\parallel )
-{\hbar^2k_\perp^2\over 2m_h} -\epsilon_g,
\end{eqnarray}
where $\lambda_c$ is the electron miniband width, $\lambda_h$ is the 
corresponding bandwidth for the holes;  $k_\parallel$ is the 
(crystal) momentum
component  parallel to the growth direction of the superlattice and
$k_\perp$ is the magnitude of the component perpendicular to the growth 
direction. The effective band thus becomes
\begin{equation}
\label{sup_dis}
\epsilon_{{\rm eff}}[\vec k] =
{1\over2}\lambda_{{\rm eff}}\cos (ak_\parallel )
+{\hbar^2k_\perp^2\over 2m_{{\rm eff}}}+\epsilon_g,
\end{equation}
where $\lambda_{{\rm eff}} = \lambda_c+\lambda_v$ is the effective
bandwidth, which again is of the same form as the original bands.
This shows that also for superlattices  
the convolution (\ref{convolution1}) leads to
an effective spectral function of the original form.

In terms of the effective spectral function the imaginary part of
the susceptibility can be written as
\begin{equation}
{\rm Im}\chi_{lm}^r(T,\omega_l ) = {d_ld_m\over 2\hbar}
\sum_{\vec k}A_{{\rm eff}}(\vec k,T,\omega_l),
\end{equation}
where we assume that the dipole matrix elements are $k$-independent.
In equilibrium the trace of the spectral function yields the density of
states for the system.
Analogously, the {\it generalized time-dependent 
density of states} (GDOS) \cite{JAU96} is defined as
\begin{equation}
\label{the_trace}
\rho (T,\omega_l) = 
{1\over\pi}\sum_{\vec k}A_{{\rm eff}}(\vec k,T,\omega_l),
\end{equation}
allowing us to write the absorption coefficient as 
\begin{equation}
\alpha_T(\omega_l)\approx {2\pi^2\omega_l |d|^2\over cn\hbar}\rho (T,\omega_l).
\end{equation}
For the remainder of this work we shall investigate the properties of
$\rho (T,\omega_l)$ for various systems.
\subsection{Gauge invariance}

To conclude this section we briefly comment on gauge invariance.
We might have, from the outset, chosen to work within
the gauge invariant formulation which has been developed in 
of high-field transport\cite{DAV88,BER91,HAU96}. Considering translationally 
invariant systems, correlation functions are made gauge invariant with the
transformation
\begin{equation}
\vec k\rightarrow \vec k+ {e\over t-t'}\int_{t'}^t{\rm d}s\vec A(s).
\label{ginvtrans}
\end{equation}
However,  the absorption coefficient
follows from  a trace operation (\ref{the_trace}),
which makes the transformation (\ref{ginvtrans}) redundant: a simple
change of variables when performing the trace
undoes (\ref{ginvtrans}) and proves that our formulation of the absorption
is gauge invariant.
\section{Parabolic bands}
\label{bulk_sec}

In this section we shall investigate the properties of the generalized
density of states for systems which can be effectively described 
by Hamiltonians yielding
parabolic bands, be it in one, two or three dimensions.
We write the effective single-band dispersion as
\begin{equation}
\epsilon [\vec k] = {\hbar^2k^2\over 2m_{{\rm eff}}} + \epsilon_g.
\end{equation}
For convenience we write $m=m_{{\rm eff}}$ and set $\epsilon_g =0$ which
shifts the energy axis such that the reference point is the band gap 
energy. We calculate the generalized density of states from
\begin{equation}
\rho^{nD} (T,\epsilon ) = 
\int_{-\infty}^\infty {\rm d}\tau e^{i\epsilon\tau/\hbar}
\rho^{nD} (T,\tau ),
\label{fourier}
\end{equation}
where
\begin{equation}
\rho^{nD} (T,\tau ) = {1\over\hbar}\int {{\rm d}^n {\vec k}\over (2\pi )^n}
\exp\Big\{
-{i\over\hbar}\int_{T-\tau/2}^{T+\tau/2}{\rm d}s
\epsilon [{\vec k}-e\vec A(s)]
\Big\}.
\end{equation}
With the vector potential (\ref{vecpot}) one obtains explicitly that
\begin{eqnarray}\nonumber
\rho^{nD} (T,\tau ) &=&
{1\over\hbar}\int {{\rm d}^n{\vec k}\over (2\pi)^n}
\exp\Bigl\{-i\big[
(\epsilon_k+\epsilon_f)\tau/\hbar \\
&&+2{e{\vec k}\cdot{\vec E}\over m\Omega^2}
\sin (\Omega T)\sin ({\Omega\tau\over 2}) \nonumber \\
&&- {\omega_f\over\Omega}\cos (2\Omega T)\sin (\Omega\tau )
\big]\Bigr\}.
\end{eqnarray}
Here $\epsilon_k = \hbar^2 k^2/2m$ and we have 
defined the fundamental energy scale
\begin{equation}
\epsilon_f = \hbar\omega_f = {e^2E^2\over 4 m\Omega^2}.
\label{omegaf}
\end{equation}
The energy $\epsilon_f$ can be interpreted classically in the following
way: consider a classical particle with charge $e$ and mass $m$ 
subjected to an electric field 
${\vec E}(t)={\vec E}\cos(\Omega t)$.  From Newton's equation of 
motion one finds that 
the mean kinetic energy of such a particle equals $\epsilon_f$.

In order to perform the Fourier-transform (\ref{fourier}) we utilize the
identity \cite{GRADSHTEYN}
\begin{equation}
\label{expansion}
\exp (ix\sin\theta ) =\sum_{n}
J_n(x)\exp (in\theta ),
\end{equation}
where $J_n(x)$ are Bessel functions; we shall henceforth write
$\sum_{n}\equiv \sum_{n = -\infty}^\infty$ to simplify the notation.
The generalized density of states becomes
\begin{eqnarray}\nonumber
\rho^{nD} (T,\epsilon ) &=&
\sum_{l,j}
\int {{\rm d}^n{\vec k}\over (2\pi )^{n-1}}
\delta [\epsilon-\epsilon_k-\epsilon_f +l\hbar\Omega ] \\
&&\times
J_{2j}\Bigl(2{e{\vec k}\cdot{\vec E}\over m\Omega^2}\sin (\Omega T) \Bigr)
\nonumber \\
&&\times
J_{l+j}\Bigl({\omega_f\over\Omega}\cos (2\Omega T)\Bigr)\;,
\label{BULK_GDOS}
\end{eqnarray}
The dimensionality is entirely contained in the remaining
momentum integration $\int{\rm d}^n{\vec k}/(2\pi )^{n-1}$.
We note that Eq.(\ref{BULK_GDOS})
implies a shift of the absorption edge by $\epsilon_f$. 
The term $l\hbar\Omega$ in the Dirac-delta function
gives rise to photonic side bands.
Since $J_{2l}(x)$ is an even function, the 
density of states is invariant under the transformation
${\vec E}\rightarrow -{\vec E}$, as expected.
In the following subsections we shall consider the 1D, 2D and 3D systems 
separately and show how the density of states smoothly evolves from a 
low field intensity regime into a high field intensity regime making the
non-linear effects of the THz field apparent.
\subsection{Generalized density of states, 1D}

\begin{figure}
\epsfxsize=8.5cm
\hspace{0.5cm}\epsfbox{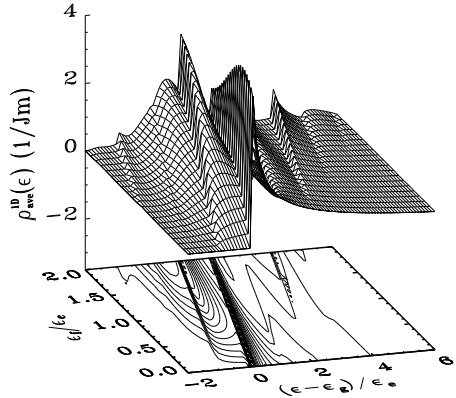}
\caption{
Time averaged generalized density of states for 1D-system 
shown for a range of FIR-intensities, $\epsilon_f/\hbar\Omega \in [0,2]$.
The band edge and the side-bands display a blue-shift,
which scales linearly with the
intensity.  Absorption extends
below the bandgap (dynamical Franz-Keldysh effect).
}
\label{1rf3}
\end{figure}

The density of states for a single-mode 1D-system
(``quantum wire'') in the absence of  external
fields is
\begin{equation}
\label{GDOS1D_1}
\rho_0^{1D}(T,\epsilon ) = {1\over\pi}\left({2 m\over\hbar^2}\right)^{1/2}
\epsilon^{-1/2}\theta (\epsilon ).
\end{equation}
In the presence of a external strong oscillating field we get from 
(\ref{BULK_GDOS}) that the GDOS is
\begin{eqnarray}\label{GDOS1D}
\rho^{1D}(T,\epsilon ) &=&
\sum_{l,j}\int_{-\infty}^\infty
{\rm d}k\delta [\epsilon-\epsilon_k-\epsilon_f+l\hbar\Omega ]\nonumber
\\
&&\times J_{2j}\left({\sqrt{32\epsilon_f\epsilon_k}\over\hbar\Omega}
\sin (\Omega T)\right)
\nonumber \\
&&\times J_{l+j}\left({\epsilon_f\over\hbar\Omega}\cos (2\Omega T)\right)
\nonumber\\
&=& 
\sum_l r^{1D}_l(T,\epsilon-\epsilon_f+l\hbar\Omega )
\rho_0^{1D}(\epsilon-\epsilon_f+l\hbar\Omega)\;,\nonumber\\
\end{eqnarray}
where the side-band weights are
\begin{eqnarray}
r^{1D}_l(T,\epsilon ) &=&
\sum_jJ_{2j}\left({\sqrt{32\epsilon_f\epsilon}\over\hbar\Omega}
\sin (\Omega T)\right)
\nonumber\\
&&\times J_{l+j}\left({\epsilon_f\over\hbar\Omega}\cos (2\Omega T)\right). 
\end{eqnarray}
We note that in the limit $\epsilon_f\rightarrow 0$
\begin{equation}
r^{1D}_l(T,\epsilon ) \rightarrow \delta_{l,0}
\end{equation}
and $\rho^{1D}(T,\epsilon ) \rightarrow \rho_0^{1D}(\epsilon )$, as expected.
If $\Omega$ is in the THz regime then most experiments would probe
the time averaged absorption. 
The time-average of the side-band weights is calculated from
\begin{eqnarray}
\bar{r}^{1D}_l(\epsilon ) &=&
\sum_j
\int_0^{2\pi}{{\rm d}s\over 2\pi}
J_{2j}\left({\sqrt{32\epsilon_f\epsilon}\over\hbar\Omega}\sin (s )\right)
\nonumber\\
&&\times J_{l+j}\left({\epsilon_f\over\hbar\Omega}\cos (2s )\right) \;,
\end{eqnarray}
which yields for $l$  odd:
\begin{eqnarray}
\bar{r}^{1D}_l(\epsilon ) &=&
\sum_j
\int_0^{2\pi}{{\rm d}s\over 2\pi}
J_{4j+2}\left({\sqrt{32\epsilon_f\epsilon}\over\hbar\Omega}\sin (s )\right)
\nonumber\\
&&\times J_{2j+l+1}\left({\epsilon_f\over\hbar\Omega}\cos (2s )\right)\;,
\end{eqnarray}
and for $l$  even
\begin{eqnarray}
\bar{r}^{1D}_l(\epsilon ) &=&
\sum_j
\int_0^{2\pi}{{\rm d}s\over 2\pi}
J_{4j}\left({\sqrt{32\epsilon_f\epsilon}\over\hbar\Omega}\sin (s )\right)
\nonumber\\
&&\times J_{2j+l}\left({\epsilon_f\over\hbar\Omega}\cos (2s )\right). 
\end{eqnarray}
At the onset of side band $l$ the side-band weight is
\begin{equation}
\bar{r}^{1D}_l(0) =
\left\{
\begin{array}{cl}
0 & \mbox{{\rm if $l$ odd}} \\
J_{l/2}^2(\epsilon_f/\hbar\Omega) & \mbox{{\rm if $l$ even,}}
\end{array}
\right.
\end{equation}
where we used the identity \cite{GRADSHTEYN}
\begin{equation}
\int_0^{2\pi}{\rm d}\theta J_{2l}(a\cos\theta) = 2\pi J_l^2(a).
\label{ident1}
\end{equation}
This shows that processes involving an odd number of photons of the THz field
are strongly suppressed. 
In Fig. \ref{1rf3} we illustrate 
$\rho_{{\rm ave}}^{1D}(\epsilon )$ for 
a range of values of $\epsilon_f/\hbar\Omega$. In the figures we write $\epsilon_e = \hbar\Omega$.
We observe all the signatures of the 
Dynamical Franz-Keldysh effect (DFK) \cite{JAU96}: Stark-like blue shift of the
main absorption-edge by $\epsilon_f$,  formation of side-bands at
$\epsilon_g+\epsilon_f\pm N\hbar\Omega$, and finite absorption within the
band-gap. 
\subsection{Generalized density of states, 2D}

Several authors have considered fields perpendicular to the quantum well, cf.
\cite{WAG96} and references in these papers; here we focus on the situation
where the electric field is in the plane of the two-dimensional electron
gas.
In such a system 
with no external field the density of states is constant,
\begin{equation}
\rho_0^{\rm 2D}(\epsilon ) = {m\over \pi\hbar^2}
\theta (\epsilon).
\label{rho0_2D}
\end{equation}
With a harmonically oscillating field we obtain from (\ref{BULK_GDOS})
\begin{eqnarray}
\label{GDOS2D_1}
\rho^{2D}(T,\epsilon ) &=&
\sum_{l,j}\int_{0}^\infty
{{\rm d}k}\,k
\int_0^{2\pi}{{\rm d}\theta\over 2\pi}
\delta [\epsilon-\epsilon_k-\epsilon_f+l\hbar\Omega ]
\nonumber\\
&&\times J_{2j}\left({\sqrt{32\epsilon_f\epsilon_k}\over\hbar\Omega}
\cos\theta\sin (\Omega T)\right)
\nonumber \\
&&\times J_{l+j}\left({\epsilon_f\over\hbar\Omega}\cos (2\Omega T)\right).
\end{eqnarray}
The integrals in (\ref{GDOS2D_1}) are again performed
using (\ref{ident1}) and writing 
the result in the side-band picture we obtain \cite{cmp_note}
\begin{equation}
\label{GDOS2D}
\rho^{2D}(T,\epsilon ) =
\sum_l r^{2D}_l(T,\epsilon-\epsilon_f+l\hbar\Omega )
\rho_0^{2D}(\epsilon-\epsilon_f+l\hbar\Omega)
\end{equation}
where the side-band weights are 
\begin{eqnarray}
r^{2D}_l(T,\epsilon ) &=&\sum_j 
J_{j}^2\left({\sqrt{32\epsilon_f\epsilon}\over\hbar\Omega}
\sin (\Omega T)\right) \\
&&\times J_{l+j}({\epsilon_f\over\hbar\Omega}\cos (2\Omega T)). \nonumber
\end{eqnarray}
Identical arguments as in the 1D case lead to
\begin{equation}
\bar{r}^{2D}_l(0) =
\left\{
\begin{array}{cl}
0 & \mbox{{\rm if $l$ odd}} \\
J_{l/2}^2(\epsilon_f/\hbar\Omega) & \mbox{{\rm if $l$ even,}}
\end{array}
\right.
\end{equation}
i.e., the same result as in the 1D-case.

\begin{figure}
\epsfxsize=8.5cm
\hspace{0.5cm}\epsfbox{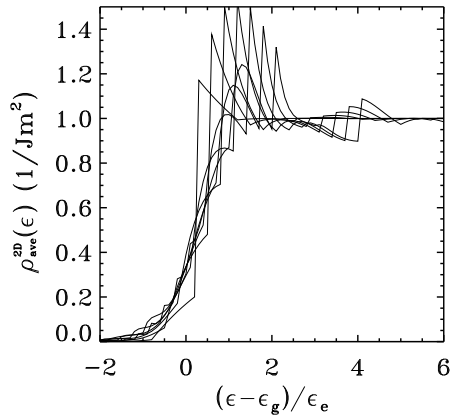}
\caption{ 
The time averaged GDOS for a 2D-system for a range of
FIR-intensities, 
$\epsilon_f/\hbar\Omega = (0.2, 0.5, 0.8, 1.1, 1.4, 1.7, 2.0)$.
At low intensities one observes a Stark-like 
blue-shift of the band edge as well
as finite absorption within the band gap. 
With increasing 
intensity side bands emerge
at $\epsilon=\epsilon_g+\epsilon_f\pm 2\hbar\Omega$.
}
\label{2rf1}
\end{figure}

\begin{figure}
\epsfxsize=8.5cm
\hspace{0.5cm}\epsfbox{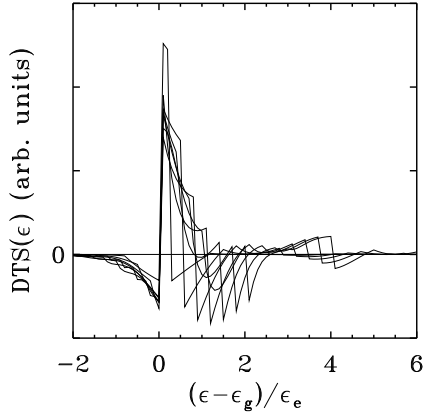}
\caption{
The DTS signal for a 2D-system 
for a range of intensities,
$\epsilon_f/\hbar\Omega = (0.2, 0.5, 0.8, 1.1, 1.4, 1.7, 2.0)$.
}
\label{2df1}
\end{figure}

As in the 1D case we have numerically investigated the time averaged GDOS 
$
\rho_{{\rm ave}}^{2D}(\epsilon ) = {\Omega\over 2\pi}\int_0^{2\pi /\Omega}
{\rm d}T\,\rho^{\rm 2D}(T,\epsilon ).
$
In Fig. \ref{2rf1} we illustrate $\rho_{{\rm ave}}^{2D}(\epsilon )$ 
for various values of $\epsilon_f/\hbar\Omega$.  Again, as in the 
1D case, we observe all the characteristics of the DFK \cite{JAU96}. 
Finally, Fig. \ref{2df1} 
shows the differential transmission (DTS) signal. 
\subsection{Generalized density of states, 3D}

\begin{figure}
\epsfxsize=8.5cm
\hspace{0.5cm}\epsfbox{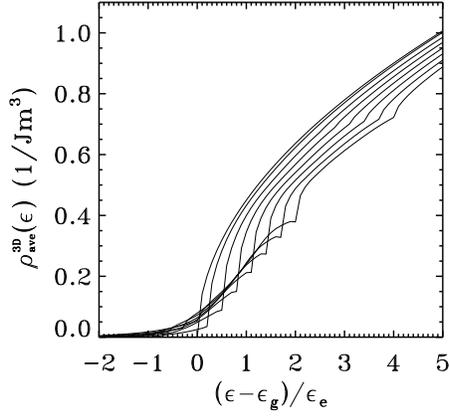}
\caption{
The time averaged generalized density of
states for a 3D-system  for a range of FIR-intensities,
$\epsilon_f/\hbar\Omega = (0.0, 0.2, 0.5, 0.8, 1.1, 1.4, 1.7, 2.0)$.
}
\label{3rf1}
\end{figure}

\begin{figure}
\epsfxsize=8.5cm
\hspace{0.5cm}\epsfbox{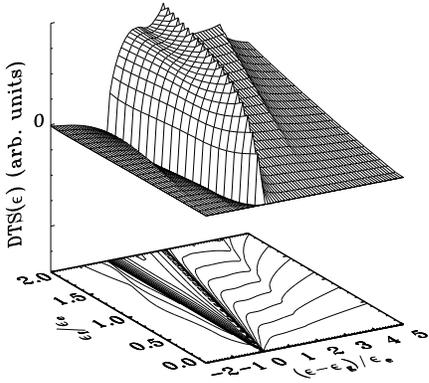}
\caption{
DTS-spectrum for a 3D-system. 
}
\label{3df3}
\end{figure}

Absorption in bulk semiconductors subject to THz radiation
was considered already long time ago by Yacoby \cite{YAC68}.
He studied transition rates between bands by investigating 
approximate solutions to the corresponding 
time-dependent Schr\"{o}dinger equation. He concluded that
transitions occur in the gap and noted reduced rates
above the gap, both in agreement with the present work. 
General quantitative results, were not presented.
The 3D field-free density of states is
\begin{equation}\nonumber
\rho^{\rm 3D}_0(\epsilon ) = {1\over 2\pi^2}\Bigl({2m\over\hbar}\Bigr)^{3/2}
\theta (\epsilon)\epsilon^{1/2}.
\end{equation}
With the external field the density of states becomes
\begin{equation}
\rho^{\rm 3D}(T,\epsilon ) = 
\sum_l r^{3D}_l(T,\epsilon-\epsilon_f+l\hbar\Omega )
\rho_0^{3D}(\epsilon-\epsilon_f+l\hbar\Omega)\;,
\end{equation}
with the side-band weights 
\begin{eqnarray}
r^{3D}_l(T,\epsilon )
&=&
\sum_j
J_{l+j}\left({\epsilon_f\over\hbar\Omega}\cos (2\Omega T)\right) 
\nonumber\\&& \times
\int_0^1{\rm d}\eta
J_{2j}\left({\sqrt{32\epsilon_f\epsilon}\over\hbar\Omega}
\sin (\Omega T)\eta \right).
\end{eqnarray}
Again, we have
\begin{equation}
\bar{r}^{3D}_l(0) =
\left\{
\begin{array}{cl}
0 & \mbox{{\rm if $l$ odd}} \\
J_{l/2}^2(\epsilon_f/\hbar\Omega) & \mbox{{\rm if $l$ even.}}
\end{array}
\right.
\end{equation}
In Fig. \ref{3rf1}
we illustrate
$\rho_{{\rm ave}}^{3D}(\epsilon )$ for various values of 
$\epsilon_f/\hbar\Omega$;
the DTS signal for the 3D case 
is shown in Fig. \ref{3df3}.

\subsection{Summary}

The main physical consequences of the THz field on linear
absorption spectrum for systems with parabolic dispersion
can be summarized as follows.
The dynamical modifications of the absorption spectrum
(i) Appear near the absorption
edge; 
(ii) They extend a few $\epsilon_e = \hbar\Omega$ around the edge, and 
(iii) They are most pronounced when $\omega_f /\Omega$ is of order unity. 

If $\Omega$ is in the THz regime,
and fields like those attainable with free electron lasers are considered
\cite{GUI93,UNT96}, then $\omega_f /\Omega\approx 1$  and the fine structure
extends over an area of several meV. Consequently, an experimental
verification of these effects should be possible.
\section{SUPERLATTICES}
\label{sup_sec}

According to the semiclassical Bloch-Boltzmann theory of transport,
a uniform electric field causes charge carriers in a periodic potential
to execute a time-periodic motion with 
frequency $\omega_B=eaE/\hbar$, where $a$ is the lattice periodicity.
Conditions for observing these Bloch oscillations are much more favorable
in superlattices than in ordinary bulk materials, and recent years have
witnessed an intense research effort culminating in the observation
of Bloch oscillations \cite{WAS93}.
In ac-fields a phenomenon called dynamical localization may occur:
if the parameter $\gamma\equiv aeE_\parallel/\hbar\Omega$ equals
a zero of $J_0$, the average velocity vanishes \cite{dynloc}.
In this section we investigate how dynamical localization 
\cite{HOL94,dynloc,ZHA94,HOL95,ROT96} manifests itself in the free particle
absorption spectra. Recently, Meier et al. \cite{MEI95}
presented results of a detailed
numerical solution of the semiconductor Bloch equations,
including excitonic effects,
and found that at dynamical localization the relative motion
exciton wave function changes from a 3D-character (i.e., localized
in $k_z$-space) to a 2D-structure (extended in $k_z$-space),
and below we shall illustrate how the same phenomenon reflects itself
in the present analytic study of free-particle properties.
\subsection{Generalized Density of States}

The starting point for our analysis is the effective dispersion 
(\ref{sup_dis}) introduced in section \ref{conv_sec} which we reproduce here 
for the convenience of the reader,
\begin{equation}
\epsilon_{{\rm eff}}[\vec k] =
{1\over2}\lambda_{{\rm eff}}\cos (ak_\parallel )
+{\hbar^2k_\perp^2\over 2m_{{\rm eff}}}+\epsilon_g.
\nonumber
\end{equation}
Henceforth we put $\epsilon_g=0$ and drop the ``eff'' subscript. We 
consider the effect of the THz field described by the vector potential 
${\vec A}(t) = -{\vec E}\sin (\Omega t)/\Omega$ where
${\vec E} = (0,0,E_\parallel )$. 
In accordance with Section \ref{gre_sec}
we calculate the generalized density of states from
\begin{eqnarray}\label{rhoslTtau}
\rho^{sl} (T,\tau ) &=& 
{m\over 2\pi^2\hbar^3}\int_0^\infty {\rm d}\epsilon_\perp\int_0^{2\pi /a}
{\rm d}k_\parallel e^{-i\epsilon_\perp\tau/\hbar}
\nonumber\\
&&\times\exp[I(T,\tau)]\;,\\
I(T,\tau)&=&-i{\lambda\over 2\hbar}
\int_{T-\tau/2}^{T+\tau/2}{\rm d}s\cos [ak_\parallel+\gamma\sin (\Omega s)].
\end{eqnarray}
We evaluate the integral
within the exponential using the identity (\ref{expansion}) with the result
\begin{eqnarray}
I(T,\tau)&=&{\lambda\over 2\hbar\Omega}
\Big\{
\cos (ak_\parallel)[{\cal C}(\Omega\tau )+J_0(\gamma ){\lambda\tau\over 2\hbar}]
\nonumber\\
&\quad&+\sin (ak_\parallel){\cal S}(\Omega\tau )
\Big\}\;,
\end{eqnarray}
where
\begin{eqnarray}
{\cal C}(z) &=& 2\sum_{n=1}^\infty
{J_{2n}(\gamma )\cos [2n\Omega T]\over n}\sin [n z] \\
{\cal S}(z) &=& 2\sum_{n=1}^\infty {J_{2n-1}(\gamma )\sin [(2n-1)\Omega T]
\over n-1/2}\nonumber \\&&\times
\sin [(n-1/2)z].
\end{eqnarray}
We have suppressed the explicit dependence of $\Omega T$ and $\gamma$ in
${\cal C}(z)$ and ${\cal S}(z)$ for simplicity. Note that both
of these functions are anti-symmetric in $z$, i.e., 
${\cal C}(-z) =-{\cal C}(z)$ and ${\cal S}(-z) =-{\cal S}(z)$. The identity
\cite{GRADSHTEYN}
\begin{equation}
\int_0^{2\pi}{{\rm d}\theta\over 2\pi}
\exp\left\{
ia\cos\theta+ib\sin\theta
\right\} = J_0(\sqrt{a^2+b^2})
\end{equation}
is the key to the next step in the evaluation of (\ref{rhoslTtau}) 
and allows us to write
\begin{equation}\label{rhoTtau}
\rho^{sl} (T,\tau )=
{m\over 2\pi^2\hbar^3a}\int_0^\infty{\rm d}
\epsilon_\perp\,e^{-i\epsilon_\perp\tau /\hbar}{\cal K}(\Omega\tau )
\end{equation}
where we have defined the kernel
\begin{equation}
\label{Kkernel}
{\cal K}(z)=
J_0\left(
{\lambda\over 2\hbar\Omega}\sqrt{ ({\cal C}(z)+J_0(\gamma )z)^2+{\cal S}^2(z) }
\right).
\end{equation}
Also here we have suppressed the explicit dependence of $\Omega T$ and 
$\gamma$. In distribution sense we can write
\begin{equation}
\int_0^\infty{\rm d}\epsilon_\perp\,e^{-i\epsilon_\perp\tau/\hbar} =
{-i\hbar\over\tau -i0^+},
\end{equation}
where $0^+$ indicates a positive infinitesimal. This expression allows us
compute the Fourier transform of (\ref{rhoTtau}):
\begin{equation}
\label{rhosl}
\rho^{sl} (T,\epsilon ) = {m\over 2\pi^2\hbar^2 a}\left(
\int_{-\infty}^\infty{\rm d}\tau{\sin (\epsilon\tau/\hbar )
\over\tau}{\cal K}(\Omega\tau )+\pi
\right).
\end{equation}
In what follows we shall examine several properties of this result.
\subsection{The field-free limit}

In the limit of vanishing field strength we have
\begin{equation}
\lim_{\gamma\rightarrow 0}{\cal K}(z) =
J_0\left({\lambda z\over 2\hbar\Omega}\right).
\label{gammalimit}
\end{equation}
Using the identity \cite{GRADSHTEYN}
\begin{equation}
\int_{-\infty}^{\infty}{{\rm d}x\over x}\,\sin (\beta x)J_0(x) =
\left\{
\begin{array}{ll}
\pi &[\beta>1]\\
2\arcsin\beta & [\beta^2<1] \\
-\pi &[\beta<-1]
\end{array}
\right.
\end{equation}
we obtain the density of states for a tight-binding superlattice,
\begin{equation}
\rho^{sl} (\epsilon ) =
{m\over\pi\hbar^2 a}\left\{
\begin{array}{ll}
1&[\epsilon>\lambda/2]\\
{1\over\pi}\arcsin (2\epsilon/\lambda )+1/2&[|\epsilon |\le\lambda/2] \\
0 &[\epsilon<-\lambda/2]
\end{array}
\right.
\end{equation}
which is familiar.
\subsection{The static limit}

In the limit  $\Omega\rightarrow 0$ one obtains
\begin{equation}
\lim_{\Omega\rightarrow 0}{\cal K}(\Omega\tau ) =
J_0\left({\lambda\over 2\hbar\omega_B}\sin (\omega_B\tau )\right).
\label{Omegalimit}
\end{equation}
Recalling the identities \cite{GRADSHTEYN}
\begin{equation}
J_0(z\sin\alpha ) = \sum_jJ_j^2(z/2)\cos (2k\alpha )
\end{equation}
and
\begin{equation}
\int_{-\infty}^\infty{\rm d}x\cos (\alpha x)\sin (\beta x)/x = \pi[
\theta (\alpha+\beta)-\theta (\alpha-\beta)
]\;,
\end{equation}
we obtain the density of states
\begin{equation}
\rho^{sl} (\epsilon) = {m\over\pi\hbar^2 a}
\sum_jJ^2_j\left({\lambda\over 4\hbar\omega_B}\right)
\theta (\epsilon + 2j\hbar\omega_B).
\label{static_rhosl}
\end{equation}
This result coincides with 
the one obtained in Refs.\cite{BLE88,VOI88}, which study both theoretically and
experimentally the effects of strong static fields on the absorption in
superlattice structures. They conclude that the step-like behavior of 
(\ref{static_rhosl}) is due localization in the growth direction
(Wannier-Stark localization).
\subsection{Dynamic localization}

As seen in the previous subsection the signature of localization in the
growth direction in a superlattice is a step-like behavior of the
density of states. This is intuitively clear since the density of states
for a 2D system (\ref{rho0_2D}) is constant. We therefore expect the
density of states to be composed of a step function for each well the
states extend into, with weight relative to the occupation in that
particular well. We shall now
show that if $J_0(\gamma ) =0$, i.e., the conditions for dynamical
localization are met, then the GDOS indeed is of this kind. 
The argument runs as follows.
If $J_0(\gamma ) =0$ then the kernel (\ref{Kkernel}) is periodic in $z$
with period $2\pi$. Furthermore, the kernel is an even function:
${\cal K}(z) = {\cal K}(-z)$. We can therefore formally write
$$
{\cal K}(\Omega\tau ) = \sum_j{\cal K}_j \cos(k\Omega\tau),
$$
which is of the same functional form as in the static limit,
Eq.(\ref{Omegalimit}).
Consequently, we
may conclude that the generalized density of states must be of the form
\begin{equation}\label{rhodynloc}
\rho^{sl}_{dyn.loc}(\epsilon ) = {m\over \pi\hbar^2 a}\sum_j{\cal K}_j
\theta (\epsilon +j\hbar\Omega),
\end{equation}
i.e., it is a superposition of step-functions. 
The weights $K_j$, however,
must be evaluated numerically, and examples are
given in the next section.
It is important to note that the ``step-length'' in the 
ac-case is determined by the frequency of the THz-field in
contrast to the static case, where it is determined by the
field strength.  The field strength enters the density of
states only through the weight-factors $K_j$.

The result (\ref{rhodynloc}) suggests that it should be possible to probe 
dynamic localization by photo-absorption: when the appropriate conditions 
are approached, the absorption coefficient should change 
qualitatively from a generic smooth behavior to a sharply defined step-like 
structure. The number of distinct steps appearing in the spectrum is determined
by the ratio $\lambda/\hbar\Omega$ which is also a measure of 
the number of wells 
the localized states span. This is fully consistent with
the results of
\cite{MEI95}, who considered a miniband of width $\lambda = 21$meV
and FIR frequency $\hbar\Omega = 20$meV, which corresponds to a single step,
and hence maximum binding energy of the corresponding
exciton which would be mostly confined to a single well.

\subsection{Numerical results}

\begin{figure}
\epsfxsize=8.5cm
\hspace{0.5cm}\epsfbox{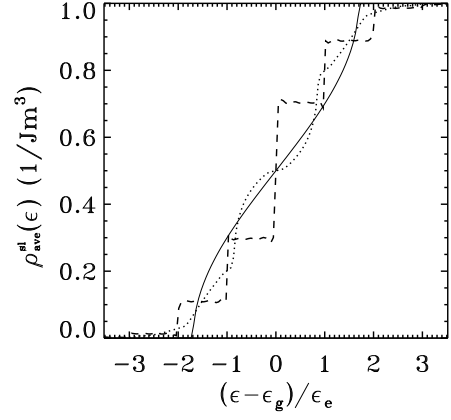}
\caption{
The time-averaged GDOS
for a superlattice with effective band-width
$\lambda = 3.4 \hbar\Omega$. 
Dots: $\gamma = 1.5$; dashes: $\gamma=2.4$.
The proximity of dynamical localization occurring
at $\gamma=2.4048...$ reflects itself in the step-wise
structure of the dashed curve.}
\label{slrf4}
\end{figure}

\begin{figure}
\epsfxsize=8.5cm
\hspace{0.5cm}\epsfbox{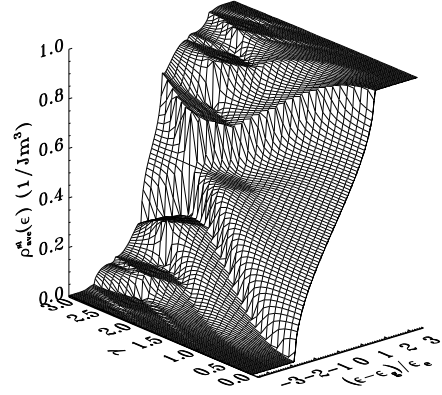}
\caption{
The time averaged GDOS
for a superlattice with $\lambda = 3.4 \hbar\Omega$
as a function of FIR-intensity.
At low $\gamma$ side bands are observed, which
merge at $\gamma = 2.4048..$
corresponding to dynamical localization. }
\label{slrf5}
\end{figure}

\begin{figure}
\epsfxsize=8.5cm
\hspace{0.5cm}\epsfbox{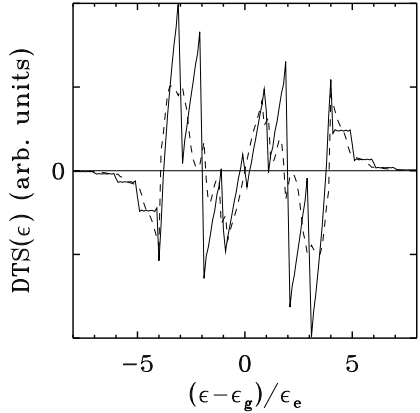}
\caption{
The differential transmission signal for superlattice 
structure with effective
mini bandwidth $\lambda = 8 \hbar\Omega$. s
Solid line: $\gamma =2.4$; dashes: 
$\gamma =2.0$. Outside the zero-field
mini-band a step-like behavior is seen, while  
inside the mini-band  DTS for dynamical localization develops
a jagged shape in contrast to the smooth behavior for the 
extended state.
}
\label{sldf2}
\end{figure}

We again focus our numerical study to the
time averaged generalized density of states 
$
\rho_{{\rm ave}}^{sl}(\epsilon ) = {\Omega\over 2\pi}\int_0^{2\pi /\Omega}
{\rm d}T\,\rho^{sl}(T,\epsilon ).
$
In Figs. \ref{slrf4} and  \ref{slrf5} we show the absorption spectra 
for a superlattice
with effective bandwidth $\lambda = 3.4\hbar\Omega$. 
The numerical results confirm the expectations of
the previous section: when the $\gamma=aeE_\parallel/\hbar\Omega$
approaches the first zero of $J_0$, which occurs at the
argument value of 2.4048.., the gradually evolving
replicas of the zero-field density of states converge
into plateaus of finite width.  The exactness of the plateaus
can be judged from Fig. \ref{slrf5}: at $\gamma=2.4048..$ the
line joining the the steps appears near vertical. 
Finally, in Fig.~\ref{sldf2} we show the DTS spectra
at dynamical localization (DL) and non-DL conditions.
There are two characteristic differences: (i) Outside
the zero-field miniband DL leads to a step-like structure
in contrast to the smooth behavior found otherwise, and
(ii) Inside the miniband the DL-spectrum distinguishes itself
by its sharp jagged structure.
\section{Conclusions}
\label{con_sec}

We have presented a theoretical formulation of linear photo-absorption
for samples under strongly non-equilibrium conditions.  Typical
non-equilibrium agents would be THz-radiation from free-electron lasers,
or ultra-short pulse measurements of transient effects.
In the present work noninteracting carriers are considered, but
the formulation allows an extension to Coulomb interactions, which
will be addressed in our future work.  Two central concepts emerge
from our analysis: a generalization of the density of states into
time-dependent conditions [GDOS defined in Eq.(\ref{the_trace})],
and photonic side-bands, which form a convenient framework for
discussing the various features of the absorption spectra.
\begin{acknowledgements}

We wish to acknowledge useful discussions with Ben Hu, Andreas Wacker
and Bj\"{o}rn Birnir. Furthermore we wish to thank Fausto Rossi
and Junichiro Kono for
valuable comments of an earlier version of the manuscript
and Kent Nordstrom for sharing details of his 
unpublished data\cite{NOR97}.
This work was partially supported by a LACOR grant no.
4157U0015-3A from Los Alamos National Laboratory.
\end{acknowledgements}
%
%


\begin{thebibliography}{32}
\bibitem{notation}
The reader should note that the symbol $\tau$ is used in a dual
meaning: it can indicate a complex-time variable on the Keldysh
contour, or the time-difference variable.  Confusion can be
avoided by noting that in the case of complex-time variables
the Green functions have a superscript $c$, while in the real-time
case they have $r,a$, or $<$.
\bibitem{cmp_note}
It is interesting to note that by comparing the GDOS in the 2D and 3D case
with Eqs. (15-18) in \cite{JAU96} one may obtain exact expansions for
the integrals in terms of Bessel functions.
\end{thebibliography}
\end{document}